# Effect of nano-size on magnetostriction of $BiFeO_3$ and exceptional magnetoelectric coupling properties of $BiFeO_3$_P(VDF-TrFE) polymer composite films for magnetic field sensor application

**Sonali Pradhan**[1,2*], **Pratik P. Deshmukh**[1], **Rahul C. Kambale**[3], **Tulshidas C. Darvade**[3,4],

**Shovan Kumar Majumder**[1] **and S. Satapathy**[1*]

[1]*Laser Biomedical Applications Division, Raja Ramanna Centre for Advanced Technology, Indore, 452 013, Madhya Pradesh, India.*

[2]*Homi Bhabha National Institute, Training School Complex, Anushakti Nagar, Mumbai, 400 094, Maharashtra, India.*

[3]*Department of Physics, Savitribai Phule Pune University, Pune, 411 007, Maharashtra, India.*

[4]*Department of Physics, Sir Parashurambhau College, Pune, 411 030, Maharashtra, India.*

**\*Address for Correspondence:**
E-mail Address: *srinu73@rrcat.gov.in;sonalipra8@gmail.com;*

**Abstract**

The existence of magnetostriction in bulk BiFeO$_3$ is still a matter of investigation and it is also an issue to investigate the magnetostriction effect in nano BiFeO$_3$. Present work demonstrates the existence of magnetostrictive strain in superparamagnetic BiFeO$_3$ nanoparticles at room temperature and the magnetoelectric coupling properties in composite form with P(VDF-TrFE). Despite few reports on the magnetostriction effect in bulk BiFeO$_3$ evidenced by the indirect method, the direct method (strain gauge) was employed in this work to examine the magnetostriction of superparamagnetic BiFeO$_3$. In addition, a high magnetoelectric coupling coefficient was observed by the lock-in technique for optimized BiFeO$_3$_P(VDF-TrFE) nanocomposite film. These nanocomposite films also exhibit room-temperature multiferroic properties. These results provide aspects of material with immense potential for practical applications in spintronics and magneto-electronics applications. We report a magnetoelectric sensor using superparamagnetic BiFeO$_3$_P(VDF-TrFE) nanocomposite film for detection of ac magnetic field.

# 1. Introduction

Magnetoelectric multiferroic materials are remarkable for strong coupling of electric, magnetic, and structural degrees of freedom, which provides ferroelectricity, ferromagnetism, and ferroelasticity simultaneously [1]. The magnetoelectric (ME) effect contributes to intensive scientific explorations in the field of sensors, actuators, memories, spintronics and transducers, which have potentially huge commercial values [2–5]. However, the magnitude and operating temperatures of observed ME coupling have been too low for practical applications. Most single-phase materials show ME coupling at extreme conditions (high magnetic field and low temperature). Therefore, the fabrication of single-phase magnetoelectric materials at room temperature is a challenge for current active research. Among several multiferroics, the only known room-temperature multiferroic material for potential practical interest is bismuth ferrite ($BiFeO_3$) which is ferroelectric ($T_C$~1100 K) and antiferromagnetic ($T_N$~ 640 K) [6]. It exhibits a weak net magnetization due to G- type magnetic ordering via Dzyaloshinskii-Moriya (D-M) interaction among nearest neighbor $Fe^{3+}$ spins with an incommensurate cycloid spin structure having a periodicity of 62 nm [7]. Bismuth ferrite shows quadratic and higher-order ME coupling at room temperature, while linear ME coupling is observed at high electric and magnetic fields [8,9]. The presence of the spin cycloid nullifies linear ME coupling between electric polarization and magnetization. However, it is important to note that bulk $BiFeO_3$ (BFO) shows negative magnetostrictive magnetoelectric coupling which was reported by *Park et al* [10]. According to the report, the magnetostrictive origin suppresses the electric polarization at the Fe site below $T_N$ outweighing the inverse D-M interaction. In 2010 *Tokunaga et al* observed field-induced polarization change with magnitudes of approximately 200 $\mu C/m^2$ along with one of the principal axes in single-domain crystals of BFO [11]. The suitability of the BFO becomes constrained in many fields because of its spiral-modulated spin structure.

Despite intense study on BFO over the past decades a fundamental understanding of structure-property correlations in BFO is still lacking, specifically the nature of the magnetic response on the size. Moreover, low dimension (< 62 nm) is expected to lead to linear ME coupling due to modification in long-range spiral modulated spin structure [12,13]. To overcome this hindrance, we have considered low dimensional confinement. Moreover, BFO displays novel physical properties with a decrease in size due to an increased surface-to-volume ratio [14]. Recently, it has been shown that BFO nanoparticles exhibit strong size-dependent magnetic properties: (1) suppression of the spiral spin structure increases with decreasing nanoparticle size, (2) uncompensated spins with spin pinning and strain anisotropies at the surface and (3) presence of oxygen vacancies and impurities [15]. BFO below a critical size affords single-domain magnetic nanoparticles exhibiting superparamagnetic (SPM) behavior. In the current work, we studied the magnetic properties of BFO nanopowders with a single domain, obtained by the auto combustion technique. The existence of magnetostrictive properties in superparamagnetic BFO is a matter of investigation. Therefore, a study on magnetostrictive properties of super paramagnet BFO has been highlighted here. Further, the use of magnetostrictive properties of superparamagnetic BFO has been explored in ferroelectric polymer composite films.

To utilize the superparamagnetic BFO nanoparticles for high performance ME coupling effect, a perfect flexible ferroelectric matrix is required. In these circumstances, P(VDF-TrFE) (poly(vinylidene fluoride-trifluoroethylene)) copolymer is an ultimate candidate for ferroelectric matrix because of its high energy density, high insulating property and good piezoelectric properties [16,17]. Moreover, among the five crystalline phases of PVDF, the β-phase is the best ferroelectric phase to implement [18]. Among the ME composites, polymer-based composites have advantages over ceramic-based composites because of non-deterioration during operation and are compatible with industrial requirements without large leakage current [19]. Nevertheless, the ME coupling effect of superparamagnetic BFO embedded in the P(VDF-TrFE) matrix has not yet been reported. In the current work, the structural, magnetic and magnetostrictive properties of superparamagnetic BFO and magnetoelectric coupling properties in BFO_P(VDF-TrFE) nanocomposite films are investigated.

In general, the ME coupling in composite systems appears due to elastic interaction between the ferroelectric and magnetic phases [20,21]. The direct ME coupling in magnetoelectric composite has been noticed to happen mainly through strain [22]. The strain induced in the magnetic phase by an external magnetic field due to magnetostriction. Magnetostriction is defined as a change in dimensions of the material in regards to an external applied magnetic field [23,24]. It is computed as $\lambda = \Delta l/l$ [25,26]. The magnetostriction of a material can be examined by direct and indirect methods. In the direct method, magnetostrictive strain is measured as a function of the externally applied DC magnetic field, which is employed here.

Due to the superparamagnetic behavior of BFO nanoparticles, the magnetic moment of nanoparticles can be more easily flipped in the polymer under an applied magnetic field at room temperature, which might create strain in the polymer matrix. Moreover, the interface effect between nanoparticles and the polymer matrix is a prominent factor for ME coupling properties, which was easily achieved through small-size BFO nanoparticles due to the increasing ratio of the interface area to volume. According to surface elasticity theory, it was found that the interfacial stress due to the inclusion of nanoparticles in polymer shows a short-range effect, which introduces internal stresses in the matrix resulting in output voltage [27,28]. Different volume % of superparamagnetic BFO nanoparticles in polymer matrix influence the dielectric, ferroelectric and magnetic properties of the nanocomposite films. Therefore, BFO_P(VDF-TrFE) nanocomposite films with different volume % (0.2, 0.5, 1, 1.5, 3 and 5%) of BFO nanoparticles were prepared to examine the room temperature multiferroic properties as well as the magnetostrictive properties of nanocomposite films and generation of magnetoelectric voltage in response to external DC magnetic field.

## 2. Experimental details
### 2.1 Sample preparation

Commercially available bismuth nitrate pentahydrate (Bi $(NO_3)_3 \cdot 5H_2O$) (Alfa Aesar, 99.99 %) and iron nitrate nonahydrate (Fe $(NO_3)_3 \cdot 9H_2O$) (Alfa Aesar, 99.99%) were used as starting materials for $BiFeO_3$ nanoparticles synthesis. They were dissolved in distilled water and stirred at room temperature. After half an hour, these two solutions were mixed and stirred at 80°C followed by the addition of glycine. Finally, the mixture was combusted at an optimized temperature of 200°C. After the self-propagating combustion, the dried gel was burnt out completely to obtain the $BiFeO_3$ nanopowders. For nanocomposite film preparation, P(VDF-TrFE) (70:30, PolyK Technologies) powder was first dissolved in N, N-dimethyl formamide (DMF) (Fluka, 99.9%) up to getting the transparent solution. Previously synthesized $BiFeO_3$ nanoparticles were added to this solution and stirred for 2 hrs to obtain a homogeneous mixture. In this method, the DMF+P(VDF-TrFE)+BFO solution was sonicated for 48 hours using the ultrasonic probe to avoid the agglomeration of nanoparticles. Nanocomposite films were prepared by using the doctor blade method [29]. Average thicknesses of 50 µm films were prepared by keeping constant velocity and gap size between the blade and glass substrate. The stretched film on the glass substrate was kept inside an oven at 100°C for polymer crystallization and solvent evaporation. Finally, the samples were cooled at room temperature overnight and peeled off from glass substrate with the help of water. The content of the nanoparticles in the polymer matrix varied as volume % of 0.2, 0.5, 1, 1.5, 3 and 5 (Inset of Fig.1 (b)).

## 2.2 Phase analysis

The compositional and structural characterizations of $BiFeO_3$ nanoparticles and composite films were obtained using X-ray diffraction (XRD) measurements performed with an X-ray diffractometer (Rigaku Geigerflex) with $CuK_\alpha 1$ (wavelength λ = 1.54 Å) as the radiation source. The XRD scans were performed with a 2θ step interval of 0.01°. Fig1(a) shows the Rietveld refinement of the XRD patterns for $BiFeO_3$ nanoparticles using Fullprof software. The well-fitted refinement ($\chi^2$=1.87) confirms a rhombohedral perovskite structure with space group symmetry *R3c* where the lattice constants value a and c are found to be 5.58Å and 13.86Å respectively (table 1). It is seen that all the nanoparticles show a single- phase and no other secondary phase is found in the refinement. The average crystallite size of $BiFeO_3$ nanoparticles is estimated to be approximately 18 nm with the help of the Scherer formula

$$P = \frac{0.9 \times \lambda}{\beta \times \cos\theta} \qquad (1)$$

Where P is the crystallite size, β is the full-width at half maxima of XRD peak in radians, λ is the X-ray wavelength and θ is the Bragg angle. The XRD patterns of all $BiFeO_3$_P(VDF-TrFE) nanocomposite films (Fig1 (b)) show that the two phases retain their identities without affecting each other.

## 2.3 Size and morphological analysis

Fig.1 (c) and (d) show Transmission electron microscopy (TEM) images of $BiFeO_3$ nanoparticles. For TEM, samples were prepared by dispersing the BFO nanoparticles in methanol using sonicator and put on carbon coated copper grid. Uniform spherical shaped nanoparticles is observed in the TEM micrograph. Field emission scanning electron microscopy (FESEM) micrograph of pure P(VDF-TrFE) and 1.5 vol % nanocomposite film are shown in Fig.1 (e) and (f) respectively. Pure P(VDF-TrFE) film shows fibrillar like crystallite β-phase of polymer. In the case of 1.5 vol % nanocomposite film, the spherulite chain like structure of the polymer is also visible with small disruption due to loading of nanoparticles. Moreover, $BiFeO_3$ nanoparticles completely embedded inside the polymer were also observed.

| Structural parameters of $BiFeO_3$ nanoparticles | Values obtained from Rietveld Refinement |
|---|---|
| Space group | R 3 c (161) |
| Lattice type | R |
| Structure | Rhombohedral |
| a = b (Å) | 5.5823 |
| c (Å) | 13.8688 |
| V(Å³) | 374.279375 |
| α = β (degrees) | 90.00 |
| γ (degrees) | 120.00 |
| Bi (x,y,z) | (0.000, 0.000, 0.400) |
| Fe (x,y,z) | (0.000, 0.000, 0.100) |
| O (x,y,z) | ( -13.8730, -47.6720, 3.7000) |
| $d_{<Bi-O>}$ (Å) | 2.6000 |
| $d_{<Fe-O>}$ (Å) | 2.0000 |
| $R_p$ (%) | 16.9 |
| $R_{wp}$ (%) | 15.9 |
| $R_{exp}$ (%) | 11.65 |
| $\chi^2$ | 1.87 |

**Table 1.** *Crystal structure of $BiFeO_3$ generated from the Rietveld refinement of XRD data using VESTA software.*

### 2.4 Dielectric studies

The dielectric behaviors of ME composite films have a great impact on their particular applications for sensors, actuators, and communications [30,31]. The room temperature dielectric studies of nanocomposite films in variation with frequency over the range of 100 Hz to 1 MHz are presented in Fig.2 (a). The real part of the dielectric constant is observed to be almost the same for 0.2, 0.5, 1 and 1.5 volume % of $BiFeO_3$_P(VDF-TrFE) nanocomposite films which are higher the value than the pure P(VDF-TrFE) film which arose from the relatively higher dielectric constant of $BiFeO_3$ nanoparticles. This suggests that composites with optimum content of nanoparticles can be used for capacitive energy devices.

In the low frequency (<1kHz) region, there is an increase in dielectric constant which is considered to be due to the existence of interfacial polarization suggested by Maxwell–

Wagnor-Sillar (MWS) and space charge polarization. The space charge polarization works in low frequencies because relaxation time is found to be larger. The interfacial polarization arises due to the misorientation of molecular chains between the amorphous region and the crystalline regions of the P(VDF-TrFE) [32]. As the frequency increases, the dielectric constant falls and remains constant which is because interfacial polarization can't follow the rapid changes of the applied field at the high frequency.

The inset of fig. 2(a) shows a tangent loss (tanδ) for nanocomposite films. In the lower frequency region, the tanδ loss for all composite samples increases concerning BFO nanoparticle content in the polymer which is related to an increase in DC conductivity and interfacial polarization. However, the tanδ values of composite films are lower than the pure P(VDF-TrFE) film except for 5 volume % nanocomposite film due to a decrease in polymer chain movement on the addition of a small percentage of BFO. On other hand, in the higher frequency region, tanδ values for all the nanocomposite films are higher than the pure P(VDF-TrFE) films for dipolar relaxation of polar C-F bonds in the polymer nanocomposite film.

The dielectric studies also allow for the determination of frequency-dependent conductivity of nanocomposite films. The electrical conductivity spectrum is shown in fig. 2b consists of two regions: low-frequency plateau (conductivity independent of frequency) and high-frequency dispersion regions (conductivity increases with frequency) [33]. The higher frequency region generally follows Johnscher's universal power law

$$\sigma_{AC} = A\omega^S \qquad (2)$$

Where A is a constant parameter, ω is the angular frequency at which the conductivity was measured and S is a dimensionless fractional parameter [34]. The conductivity value increases at a higher frequency region (although the overall increase is very low) from $6.72\times10^{-7}$ S cm$^{-1}$

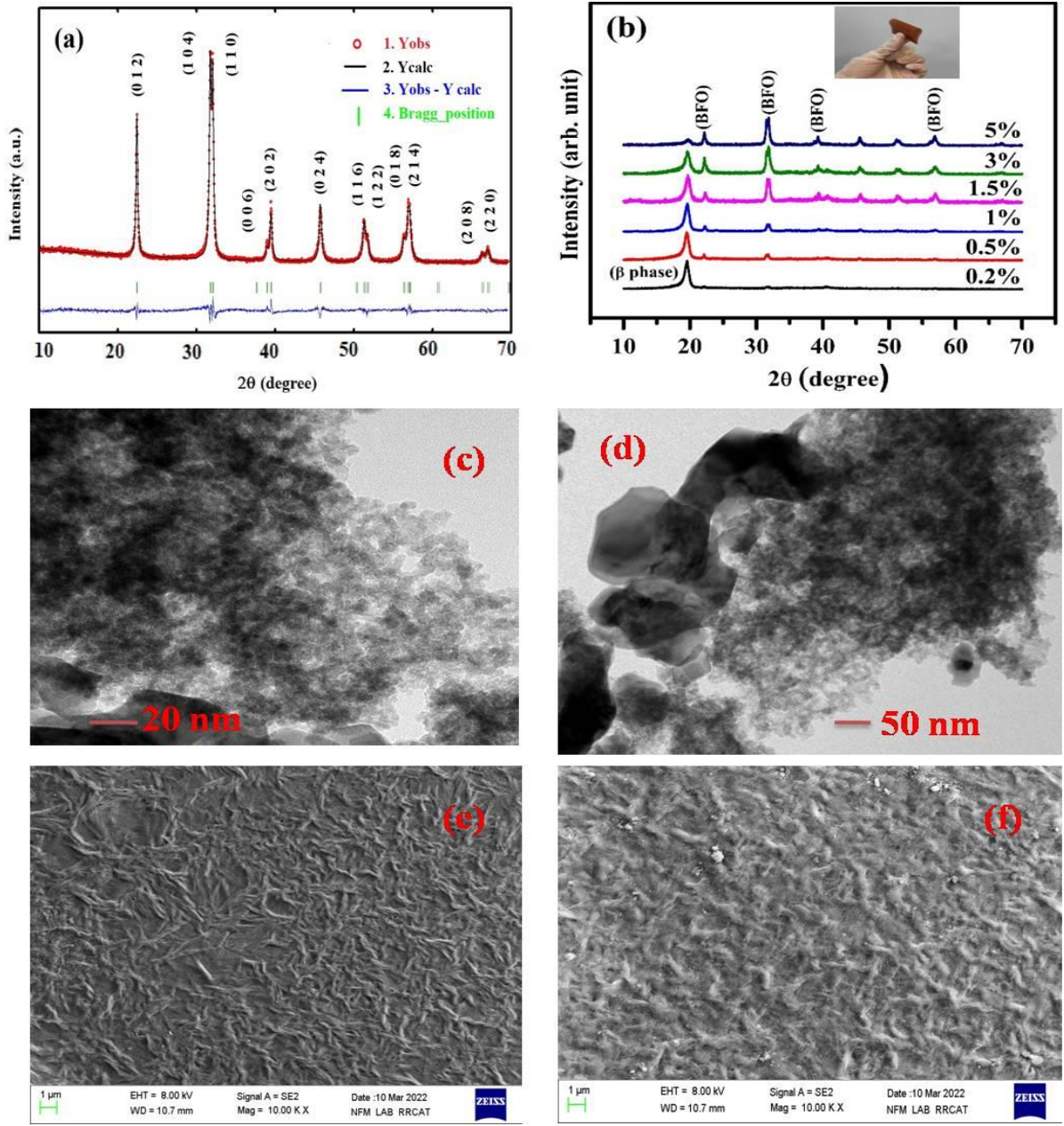

**Fig.1** *(a) Rietveld refinement of BiFeO$_3$ nanoparticles; (b) XRD pattern of BiFeO$_3$ _ P(VDF-TrFE) nanocomposite films (inset: photographic image of 1.5 vol % nanocomposite film); (c) and (d) TEM images of BiFeO$_3$ nanoparticles; FESEM images of (e) pure P(VDF-TrFE) and (f) 1.5 vol % BiFeO$_3$ _ P(VDF-TrFE) nanocomposite film*

to 3.44×10$^{-6}$ S cm$^{-1}$ with an increase of BFO nanoparticles from 0.2 to 5-volume % in polymer. This is because of dipolar relaxation of the permanent dipole in a short time. The value of the exponent S which is frequency-dependent was determined to be 0.997 ± 0.0002 which is given in the inset of Fig 2(b).

## 2.5 Ferroelectric properties

To observe the ferroelectric properties of BiFeO$_3$_P(VDF-TrFE) nanocomposite films, the room temperature polarization versus electric field (P-E) measurements have been performed at 50 Hz using the P–E loop tracer (Marine India). All nanocomposite films (except 3 and 5 vol %) show well-saturated hysteresis loops where for higher vol % of BFO nanoparticles in polymer the loops become lossy [supplementary file, Fig. S1]. Pure P(VDF-TrFE) film shows remnant polarization of about 2.04 µC/cm$^2$ at coercive field 480 kV/cm (Fig. 2(c)). Increasing BFO nanoparticle content (up to 1.5 vol %) in the P(VDF-TrFE) will enhance the polarization due to the introduction of additional free charges for the stabilization of the polarization domain [35]. However, up to certain content of BFO nanoparticles in the polymer causes a maximum polarization, which may be attributed due to an increase in interfacial areas [36]. In our case, the maximum remnant polarization was observed to be 2.74 µC/cm$^2$ for 1.5 vol % BiFeO$_3$_P(VDF-TrFE) nanocomposite film corresponding to coercive field 410 kV/cm Fig 2(d).

## 2.6 Magnetic characterizations

The magnetic characterizations of BFO nanoparticles and BFO_P(VDF-TrFE) nanocomposite films were carried out using S700X SQUID magnetometer (Cryogenics Ltd, UK). Fig 3 (a) shows the isothermal magnetization hysteresis loops (*M* vs. *H*) for the BFO nanoparticles measured at room temperature. We inferred an *S*-shape M-H loop with null coercive field ($H_C$) indicating superparamagnetic-like behavior. In fact, at higher fields, the magnetization does not reach saturation. This might be due to disordered magnetic moments located at the nanoparticle surface. To explore the superparamagnetic behavior in BFO nanoparticles the temperature-dependent zero-field cooling (ZFC) and field cooling (FC) magnetization were performed at 300 Oe. The FC magnetization monotonically decreases, while the ZFC curve identified at three different ranges of temperatures (Fig. 3(b)), indicated in blue, green, and cyan). The ZFC curve exhibits a narrow peak around 50K which can be ascribed to a blocking temperature ($T_B$), *i.e.*, this temperature is closer to the system which undergoes a blocking-to-unblocking transition. Above the blocking temperature (at high T) the thermal energy is much larger than the anisotropy energy. So, the magnetic moments of the particles fluctuate freely with temperature and a paramagnetic-like behavior is observed and the particles are in the superparamagnetic (SPM) state. On the other hand, at very low temperatures the thermal energy is not enough to switch the magnetic moment, which results in the particle moment to confine along the anisotropy direction. In this situation, the particles are said to be in a blocked state, which is confirmed by the sharp fall in the ZFC curve in the low-temperature region. It has been observed that the ZFC and FC curves merge at a temperature above the maximum reached by the ZFC curve. This temperature is known as the irreversibility temperature ($T_i$). In fact, for mono-size non-interacting magnetic nanoparticles, the blocking temperature can be considered the inflection point of the ZFC curve. In our case, there is a size distribution, which is confirmed by SEM analysis and from a wide blocking region in the ZFC curve. When magnetic nanoparticles of different sizes are considered, the maximum of the ZFC magnetization curve is influenced by both the size dispersion and the mean particle size. This indicates the distribution of blocking temperature $f(T_B)$ associated with a nanoparticles size distribution because different particles are associated with different energy barriers leading to different $T_B$ for each size fraction [2,37]. The blocking temperature distribution $f(T_B)$ is estimated using the reduced-magnetization derivative $\frac{d(M_{ZFC} - M_{FC})}{dT}$ [38]. Fig. 3 (c) shows well fitted lognormal-type function of $f(T_B)$ distributions results to $<T_B>$ = 16.4 K ($\sigma$ = 0.60) for BFO nanoparticles. Fig. 3(d) shows a well saturated M-H curve of all BFO_P(VDF-TrFE) nanocomposite films carried out at room temperature. The magnetization of nanocomposite films was observed to be gradually increasing with an increase in vol % of BFO in a polymer. All the nanocomposite films have zero retentivity and coercivity with $M_S$ value 0.28, 0.35, 0.41, 0.46, 0.85 and 1.26 emu/cc for 0.2, 0.5, 1, 1.5, 3 and 5 vol % respectively. Moreover, there is a distinction observed in the ZFC-FC magnetization curve of low (0.2 %) and high vol % (1.5 %) of BFO in BFO_P(VDF-TrFE) nanocomposite film which is shown in Fig. 3(e) and (f). In 0.2 vol % of the nanocomposite, the ZFC curve exhibits a broad maximum in the lower temperature region whereas a relatively narrow peak was obtained in the case of 1.5 vol % of the nanocomposite. The variation in the ZFC curve for different vol % was studied experimentally and theoretically which was attributed to interparticle interaction in magnetic granular systems [39]. Actually, for a highly dense system, the determination of blocking temperature is very difficult due to the interaction of particles with neighbors. Nevertheless, it should be noticed that when nanoparticles are dispersed in a polymer the interaction among the particles becomes negligible [40]. Because of the non-interacting superparamagnetic model, we fitted the derivative $\frac{d(M_{ZFC}-M_{FC})}{dT}$ of experimental data with log normal distribution function to get blocking temperature distribution of nanoparticles in polymer matrix. The experimental and fitted curve well agreed with each other satisfying non-interacting super paramagnetic model (inset of fig 3(e) and (f)). We obtained mean blocking temperature $<T_B>$ = 17.4 K ($\sigma$ =0.46) for 0.2 vol % and $<T_B>$ = 10.9 K ($\sigma$ =0.66) for 1.5 vol % nanocomposite films. Hence, when the nanoparticles are dispersed on polymer matrix, blocking temperature decreases with the nanoparticle's concentration.

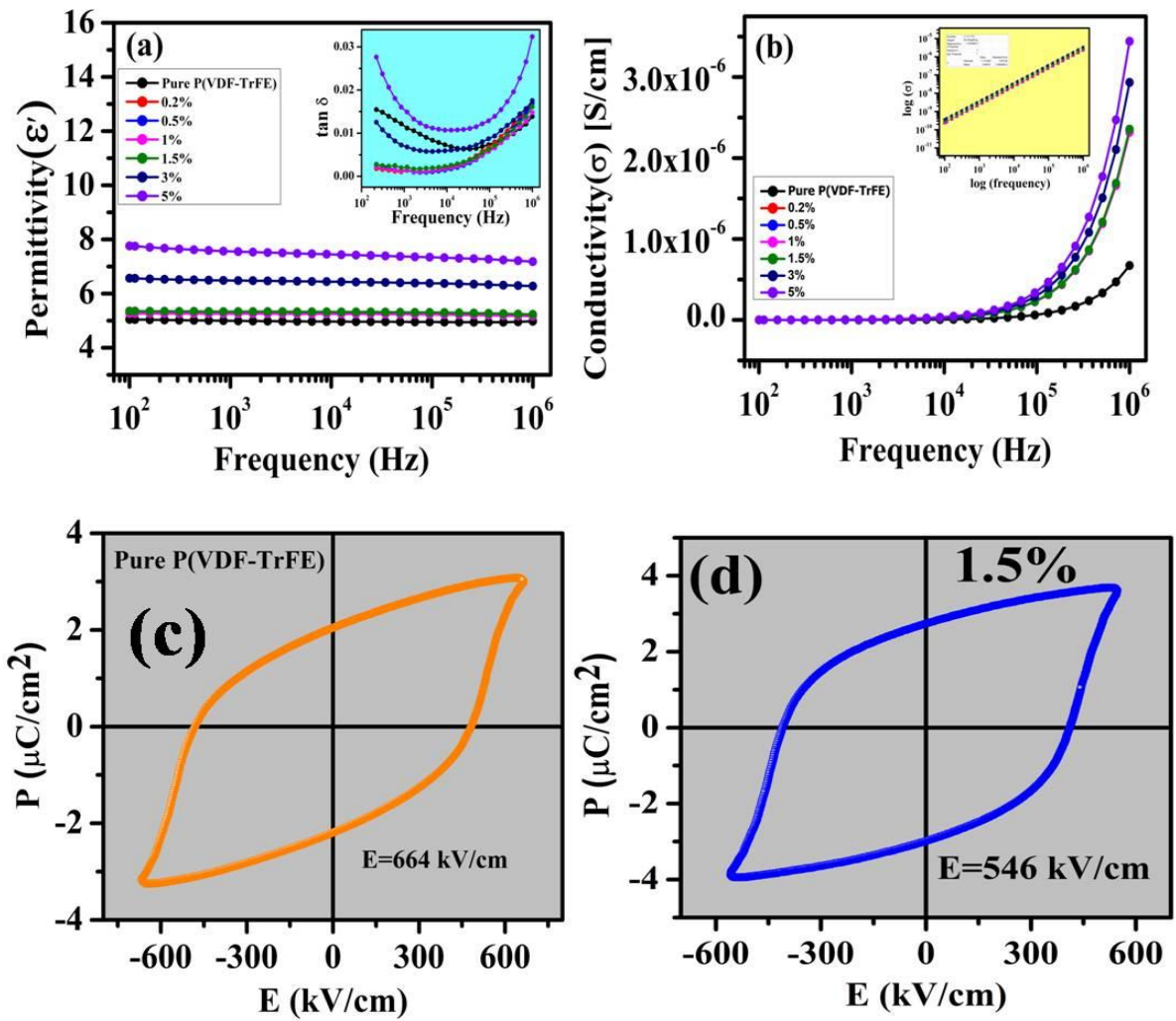

***Fig.2.*** *Frequency dependent (a) dielectric permittivity and (inset: loss tangent versus frequency) and (b) conductivity versus frequency for pure P(VDF-TrFE) film and BiFeO$_3$_P(VDF-TrFE) nanocomposite films; Ferroelectric polarisation versus electric field curve for (c) pure P(VDF-TrFE) film and (d) 1.5 vol % BiFeO$_3$_P(VDF-TrFE)*

**2.7 Strain-mediated ME coupling**

P(VDF-TrFE ) and BFO nanoparticles are expected to the couple via elastic interactions at the interfaces due to their piezoelectric and magnetostrictive characteristics, respectively. It is important to note that bulk BFO does not have a net magnetization due to its antiferromagnetic nature while at the nanoscale the BFO shows weak ferromagnetism owing to suppression of spin spiral periodicity. In the attempt to investigate the coupling between the electric and magnetic order parameters in P(VDF-TrFE ) and BFO nanoparticles, the ME coupling of BFO_P(VDF-TrFE) nanocomposite film was carried out at room temperature using the lock-in amplifier method (supplied by Marine India). In polymer composite systems, the ME coupling arises due to elastic interactions among ferroelectric and magnetic domains via magnetostriction [22,41]. In our case, due to modification in long-range spiral modulated spin structure of BFO nanoparticles lattice strain is developed. In addition to that, surface properties increase because of the nano-size effect of BFO nanoparticles. It is important to note that interface has become a prominent factor for ME coupling in polymer composite systems.

Therefore, the nano-size effect and developed lattice strain due to suppression of spin cycloid structure have become a source to anticipate ME coupling in BFO_P(VDF-TrFE) nanocomposite film. Depending on the elastic interactions and type of externally applied field (electric or magnetic), there are generally two types of ME coupling: (i) direct ME coupling and (ii) converse ME coupling. In this work, we examined direct ME coupling according to the equation [20]

$$\alpha_{ij}^H = \frac{\partial P_i}{\partial H_j} = \varepsilon_0 \varepsilon_{ii} \left( \frac{\partial E_i}{\partial H_j} \right) = \varepsilon_0 \epsilon_r \alpha_V^H \qquad (3)$$

Where $\qquad \alpha_V^H = \frac{1}{t} \left( \frac{\partial V}{\partial H} \right) \qquad (4)$

The ME voltage coefficient $\alpha_V^H$ was evaluated using the formula (4) where, $\partial V$ is the output voltage, $t$ is the thickness of the nanocomposite film and $\partial H$ is the applied AC magnetic field. Before the ME measurement, the sample was electrically poled by applying an electric field of 400 kV/cm along a perpendicular direction to the thickness of the sample. An ac magnetic field $\partial H$ of 25 Oe at a frequency of 10 kHz superimposed on an applied direct current bias field to measure the change in voltage $\partial V$ across the nanocomposite film. The ME voltage coefficient versus applied DC magnetic field plot for 1.5 vol % BFO_P(VDF-TrFE) nanocomposite film was shown in Fig. 4(a). A significant transverse magnetic field induced electric voltage ($\alpha_{ME}^{31}$) of ~ 104 mV/Oe-cm was found across the thickness of 1.5 vol % BFO_P(VDF-TrFE) nanocomposite film using the above equation (4). It was observed that the ME coupling coefficient increases with an increase in DC magnetic field and reaches a maximum at 1.7 kOe followed by a decrease in value at a higher DC magnetic field. The initial enhancement of the ME coupling coefficient might be due to an increase in magnetization of BFO nanoparticles in the polymer matrix.

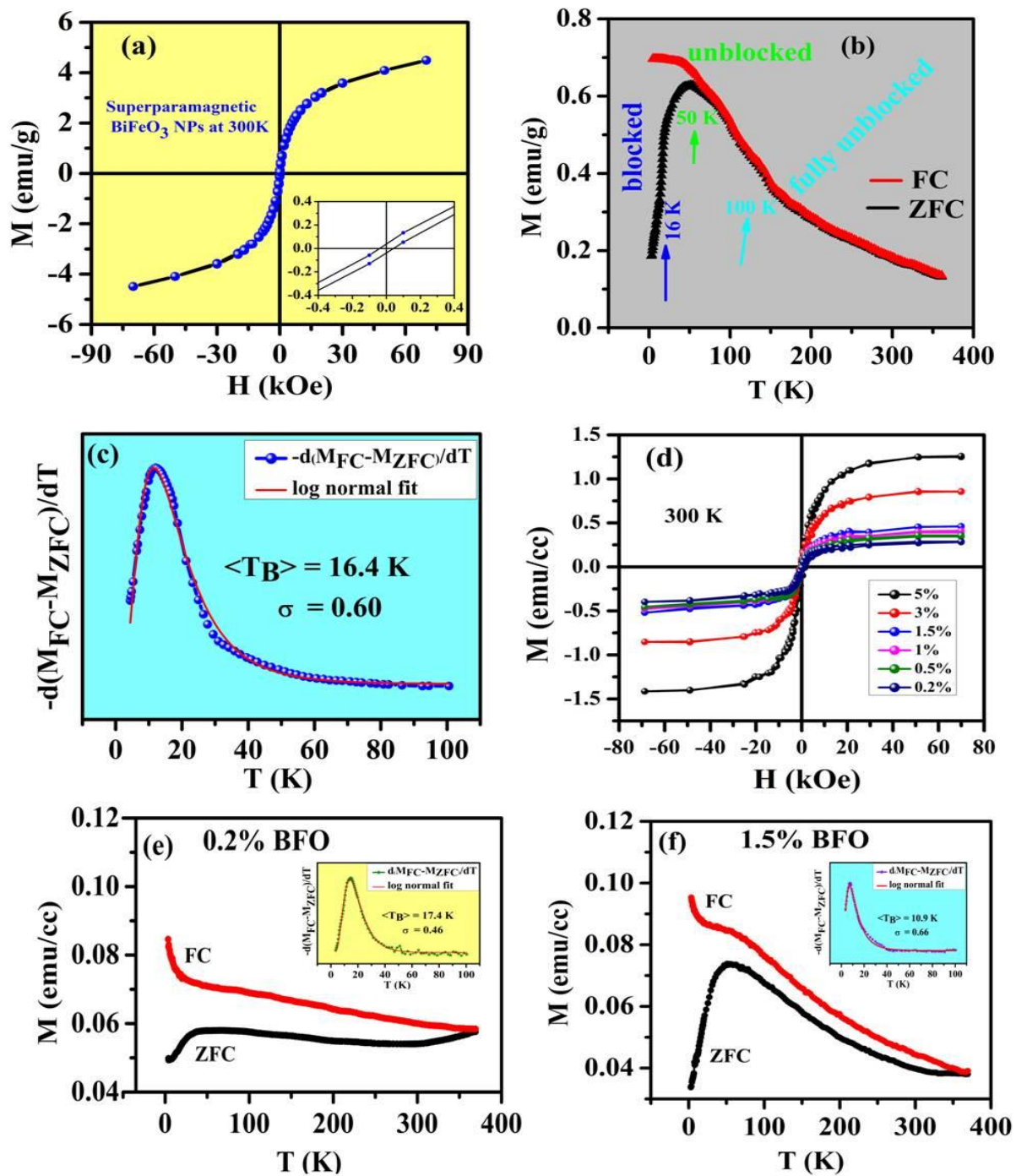

**Fig.3.***(a) M-H hysteresis curve at 300 K and (b) ZFC-FC magnetisation curve at 300 Oe magnetic field for BiFeO$_3$ nanoparticles; (c) lognormal fitting curve for derivative of $M_{ZFC}$-$M_{FC}$ with respect to temperature for BiFeO$_3$ nanoparticles; (d)M-H hysteresis curve for BiFeO$_3$_P(VDF-TrFE) nanocomposite film at room temperature; (e) and (f) ZFC-FC magnetisation curve at 300 Oe magnetic fieldfor 0.2 vol % and 1.5 vol % BiFeO$_3$_P(VDF-TrFE) nanocomposite film, respectively*

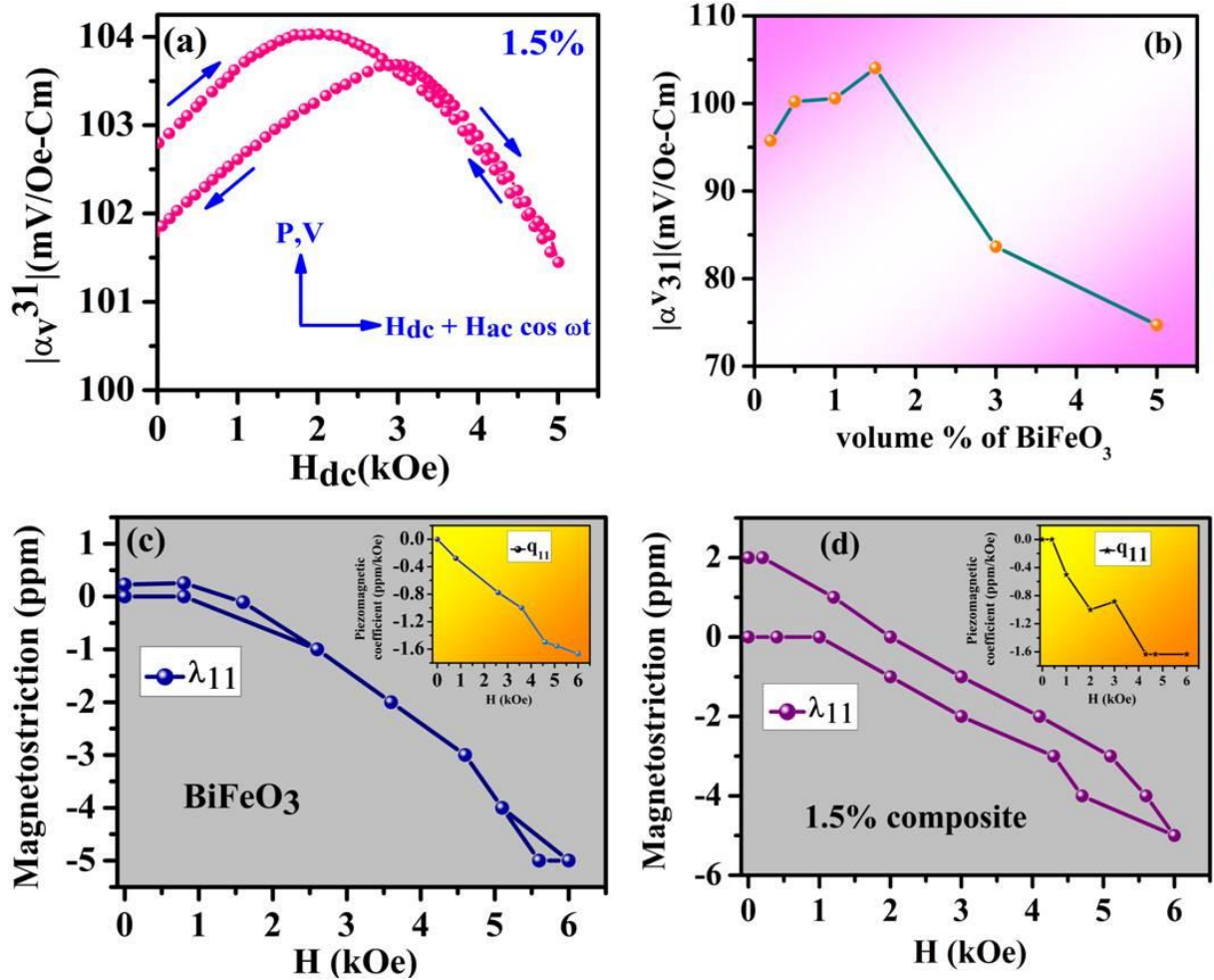

**Fig.4.** *(a) Magnetoelectric voltage coefficient (transverse) in response to DC magnetic field for 1.5 vol % BiFeO$_3$_P(VDF-TrFE) nanocomposite film; (b) variation of ME voltage coefficient with respect to volume percentage of BiFeO$_3$ nanoparticles in P(VDF-TrFE) polymer; (c) and (d) magnetostriction strain measurement with DC magnetic field for BiFeO$_3$ nanoparticles and 1.5 vol % BiFeO$_3$_P(VDF-TrFE) nanocomposite film respectively (inset: piezomagnetic coefficient ($\frac{d\lambda}{dH}$) versus DC magnetic field curve)*

After reaching a certain DC magnetic field (in this case 1.7 kOe) the ME voltage coefficient decreases for further increase of DC magnetic field due to a decrease in elastic interaction between the nanoparticles and polymer matrix. The variation of the ME voltage coefficient with a volume percentage of nanoparticles in the polymer is shown in Fig 4(b). The initial increase in the ME coupling coefficient (from 0.2 to 1.5 vol %) may be described by an increase in magnetostriction in the ferroelectric phase as well as the increase of crystallization of the electroactive $\beta$-phase due to the inclusion of ferrite nanoparticles in PVDF copolymer [42,43]. Further increase of nanoparticles in the polymer phase leads to disruption of the polymer chain structure. Because of the loading of more nanoparticles in the ferroelectric phase, the fillers cannot disperse in the polymer phase properly, which results in a fall in ME response from 1.5% to 5%. Although a few reports are available in the literature on magnetoelectric effects in superparamagnetic polymer composites. It is found that the obtained value of magnetoelectric voltage coefficient ($\alpha_{ME}$) of ~ 104 mV/Oe-cm of the present work for BFO_P(VDF-TrFE) nanocomposite is highest as compared to reported values so far (table 2).

| Superparamagnetic nanoparticles in P(VDF-TrFE) | ME coefficient (mV/Oe-cm) | Reference |
|---|---|---|
| $BiFeO_3$/P(VDF-TrFE) | 104 | In our work |
| $BiFeO_3$/PVDF | 2.7 | [44] |
| $Fe_3O_4$/P(VDF-TrFE) | 0.8 | [19] |
| $Zn_{0.2}Mn_{0.8}Fe_2O_4$/ P(VDF-TrFE) | 0.16 | [19] |
| $CoFe_2O_4$/P(VDF-TrFE) | 34 | [45] |
| $Ni_{0.5}Zn_{0.5}Fe_2O_4$/P(VDF-TrFE) | 1.35 | [46] |

**Table 2.** *Data on Magnetoelectric voltage coefficient for superparamagnetic-polymer based multiferroic system.*

When the magnetic field is applied to the composite film, mechanical strain develops in the magnetic phase via the magnetostrictive effect. The developed strain in the magnetic phase transfers to the ferroelectric phase and generates electrical voltage via the piezoelectric effect. In the present work, we are applying external magnetic field along parallel direction to the sample (i.e 1 in $\alpha_{31}$) and measuring the output voltage across the thickness of the sample or perpendicular direction of the sample (i.e 3 in $\alpha_{31}$) according to the equation

$$dP_3 = \alpha_{31} \, dH_1. \qquad (5)$$

The observed ME coefficient is governed by the equation

$$\alpha = m_V \, (dS/dH)_{BFO} \, (1-m_V) \, (g_{33}C_{33})_{polymer} \qquad (6)$$

Here $m_V$ is the volume percentage of BFO nanoparticles, dS/dH is the change in strain with magnetic field, $g_{33}$ is the piezoelectric voltage constant and $C_{33}$ is the stiffness constant. Hence this relation indicates that ME coefficient is highly dependent on the piezoelectric coefficient, volume percentage of BFO nanoparticles and magnetostriction. Due to application of an external DC magnetic field, strain is generated in BFO nanoparticles along the applied magnetic field direction (axial strain). To keep the volume of nanoparticles to be conserved, a compressive strain (transverse strain) will be developed in nanoparticles (as BFO has negative magnetostriction coefficient), which will be in perpendicular direction to the applied magnetic field. The stress corresponding to this compressive strain will transfer to the P(VDF-TrFE) polymer via its interface which will act as an external stress to the piezoelectric phase (which is in "z" direction). This external stress causes polarization in the P(VDF-TrFE) polymer due to piezoelectric effect according to the equation

$$D_i = e_{ij} \, E_j + d_{im} \, \sigma_m \qquad (7)$$

Where, D is electric displacement, e is dielectric constant, E is electric field, d is piezoelectric coefficient and $\sigma$ is stress. Since we are considering the $d_{33}$ piezoelectric coefficient of P(VDF-TrFE) (because sample is poled along direction 3), equation (7) becomes

$$D_3 = e_{3j} \, E_j + d_{33} \, \sigma_3 \qquad (8)$$

According to the above equation (8), maximum output voltage will be obtained along the direction 3 i.e perpendicular to the sample. Moreover, we are measuring the output voltage along direction 3 (according to the equation (5)) for which we are getting maximum ME coupling coefficient along perpendicular direction to the sample. In addition, the ME effect is also directly proportional to the magnetostriction according to the equation

$$\alpha^H \propto q.d \qquad (9)$$

Where d is the piezoelectric coefficient and $q = \frac{d\lambda}{dH}$ ($\lambda$ is magnetostriction) is the piezomagnetic coefficient. Therefore, we investigated the magnetostrictive strain behavior of BFO nanoparticles and nanocomposite film in response to the DC magnetic field. The $\lambda(H)$ curves obtained in a parallel configuration with an applied magnetic field are plotted in Fig. 4(c) and (d) together with the d$\lambda$/dH vs. H dependences for the BFO nanoparticles and 1.5 vol % nanocomposite sample (inset of Fig. 4(c) and (d)). The magnetostrictive strain was examined using the four-wire method and the data were recorded on a hand-held data logger TC-32K (TYPE S-2770) instrument. Magnetostriction was measured along the parallel direction to the applied magnetic field by using a 350-ohm resistive strain gauge. A small negative magnetostrictive strain (-4 ppm at 5 kOe) was observed in both nanoparticles and nanocomposite samples. It is important to note that the strain has linear behavior with applied DC magnetic field over the range 0 to 6 kOe. It should be noted here that the maximum applied magnetic field was not enough for saturation of magnetostrictive strain of superparamagnetic BFO nanoparticles. A similar trend was also observed in the ME voltage coefficient where the voltage coefficient decreases without attaining saturation up to 5 kOe magnetic field. To our knowledge, this is the first magnetostriction measurement result by the direct method reported for the superparamagnetic BFO sample.

## 2.8 Magnetoelectric sensor in BFO_P(VDF-TrFE) nanocomposite films

The magnetic field sensing properties of the composite film was further investigated by recording the output voltage generated from the sample without any amplification in an ac magnetic field in a digital storage oscilloscope (DSO). As per the result discussed in the previous section, the maximum ME coupling voltage coefficient was obtained for 1.5 vol % BFO_P(VDF-TrFE) nanocomposite film. Hence this composite film with an effective area of 1.5×1.5 cm$^2$ was taken for sensor application.

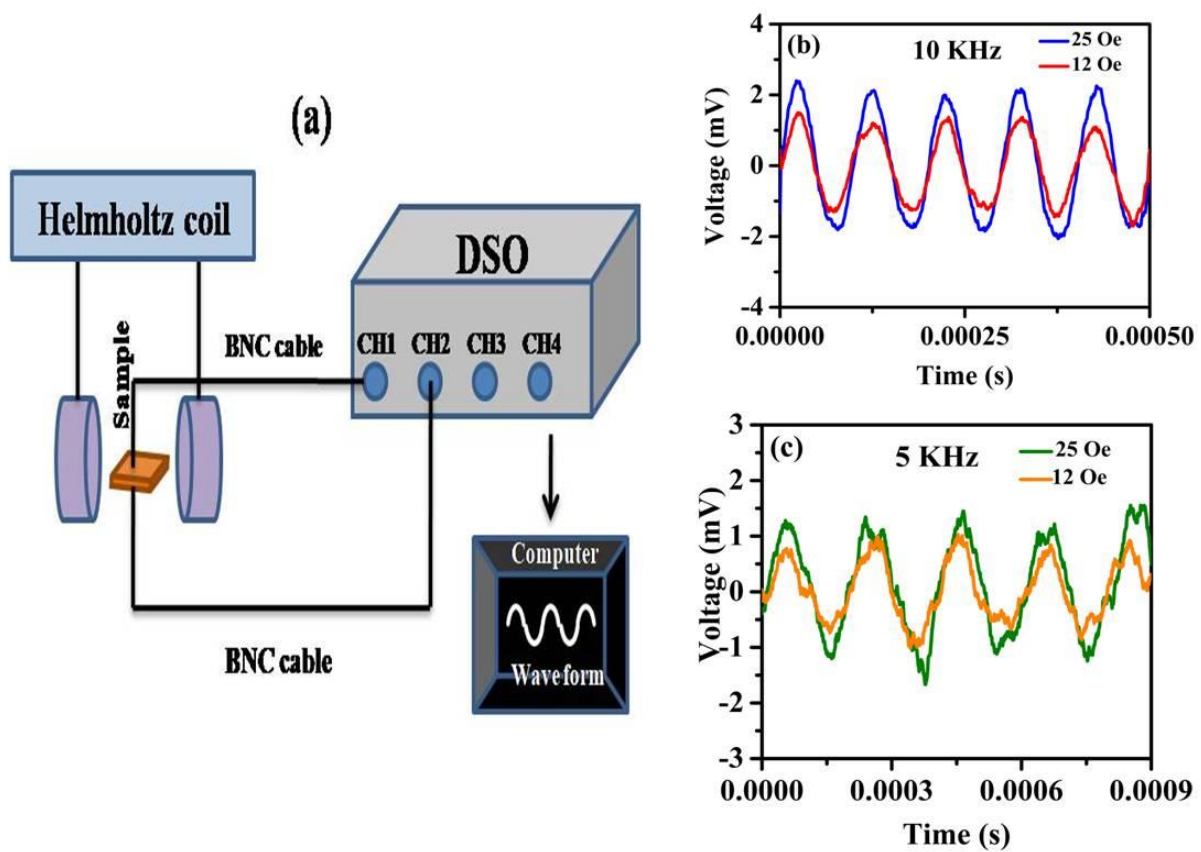

**Fig.5**. *(a) Schematic diagram of BFO_P(VDF-TrFE) ME sensor; (b) and (c) Voltage response of 1.5 vol % BFO_P(VDF-TrFE) nanocomposite film at 25 and 12 Oe AC magnetic field for 10 kHz and 5 kHz, respectively.*

First of all, the bottom and top surface of the film were coated with gold to make it conductive and after that ferroelectrically poled in a perpendicular direction across the thickness of the film. Then it was kept inside a Helmholtz coil along its axis. Two BNC cables taken from the top and bottom surface of the sample are connected to a DSO for measuring the output voltage difference between the top and bottom electrodes. Fig 5(a) shows the schematic of the above arrangement. The output voltage was recorded by applying AC magnetic fields 25 Oe and 12 Oe at two different frequencies i.e., 5 kHz and 10 kHz (Fig. 5(b) and (c)). The maximum output voltage of nearly about 2.3 mV was harvested for an AC magnetic field of 25 Oe at a frequency of 10 kHz. The ME effect is observed only due to stress-strain transfer via elastic coupling. Here, we have piezoelectric and magnetic systems having ability to get strains after application of electric and magnetic field and vice-versa. The cumulative effect is to check effective stress-strain is important in this composite and since here the magnetic phase is in nanometric range will exhibit collective behaviour during strained situation. Before the measurement, the composite film was first of all electrically poled by applying an electric field of 400 kV/cm along a perpendicular direction to its thickness. Since here the ferroelectric phase is P(VDF-TrFE), strain is induced during the application of electric field due to the converse piezoelectric effect. This piezoelectric strain is transferred to the BFO nanoparticles through the interface in BFO_P(VDF-TrFE) nanocomposite film. Hence, there is possibility that the transferred strain freezes the magnetic moment direction of the BFO nanoparticles which are in direct contact with P(VDF-TrFE) [47–49]. When AC magnetic field was applied to this nanocomposite film using a Helmholtz coil, the magnetic moment aligned along a direction produced output voltage by direct ME effect due to the magnetostrictive property of the BFO nanoparticles.

## 3. Conclusions

In summary, the self-biased ME sensor was fabricated in a two-step process. In the first step, magnetoelectric BFO superparamagnetic nanoparticles were synthesized using the auto combustion method. In the second, 0-3 BFO_P(VDF-TrFE) nanocomposite films were prepared by incorporating different vol % of BFO nanoparticles in P(VDF-TrFE) using the solution casting method. These BFO_P(VDF-TrFE) nanocomposite films exhibit well saturated ferroelectric loop and low dielectric loss (tan$\delta$ < 0.005 for 1.5 vol %) which make these nanocomposite films a strong candidate for practical device applications. Unlike bulk BFO, a negative magnetostrictive strain (~4 ppm at 5 kOe) was generated in both nanoparticles and nanocomposite samples. Therefore, a room temperature ME coupling effect ($\alpha_{ME}^{31}$ ~ 104 mV/Oe-cm) was observed in BFO_P(VDF-TrFE) nanocomposite films for which those nanocomposite films can be employed for practical applications like magnetic field sensors, data storage and switching devices. The direct output voltage of 2.3 mV was observed by applying an AC magnetic field of ~ 25 Oe at a frequency of 10 kHz to 1.5 vol % nanocomposite film in magnetic sensor geometry.

**Acknowledgments**

The authors thank RRCAT and HBNI, Mumbai (Sanction No. DAE/LBAD/5401-00-206-83-00-52/-, LT830006) for financial support. The authors acknowledge Shri Prem Kumar and Mrs. Rashmi Singh (SO/G) for their help in XRD and FESEM measurements, respectively. The authors gratefully acknowledge Dr. Azam Ali Khan of IIT Bombay for providing the TEM data. Dr. Rahul C. Kambale thankfully acknowledges UGC-DAE CSR Indore, Government of India (Ref.CSR-IC-TIMR-07/CRS-274/2017-18/1280, Collaborative Research Scheme.) for providing the research funds to carry out this research work.

# References

[1]  L.E. Fuentes-Cobas, J.A. Matutes-Aquino and M.E. Fuentes-Montero, Materials Research 19 (2011).

[2]  Moscoso-Londoño O, Muraca D, Pirota K R, Knobel M, Tancredi P, Socolovsky L M, Mendoza Zélis P, Coral D, Fernández van Raap M B, Wolff U, Neu V, Damm C and de Oliveira C L P , Different approaches to analyze the dipolar interaction effects on diluted and concentrated granular superparamagnetic systems, J. Magn. Magn. Mater. 428 (2017) 105–18.

[3] B.G. Catalan, J.F. Scott, Physics and Applications of Bismuth Ferrite, Adv. Mater. 21 (2009) 2463–2485.

[4] Jang H M, Han H and Lee J H, Spin-coupling-induced Improper Polarizations and Latent Magnetization in Multiferroic BiFeO3, Sci. Rep. 8 (2018) 1–14.

[5] R. Ramesh, Nicola A Spaldin, Multiferroics : progress and prospects in thin films, R. Article. 3 (2009) 20–28.

[6] J.T. Zhang, X.M. Lu, J. Zhou, H. Sun, J. Su, C.C. Ju, F.Z. Huang, J.S. Zhu, Origin of magnetic anisotropy and spiral spin order in multiferroic BiFeO3, Appl. Phys. Lett. 100 (2012) 242413.

[7] J.G. Park, M.D. Le, J. Jeong, S. Lee, Structure and spin dynamics of multiferroic BiFeO3, J. Phys. Condens. Matter.26 (2014) 433202.

[8] Kawachi S, Miyahara S, Ito T, Miyake A, Furukawa N, Yamaura J I and Tokunaga M, Direct coupling of ferromagnetic moment and ferroelectric polarization in BiFeO3, Phys. Rev. B 100 (2019) 140412.

[9] Tokunaga M, Akaki M, Ito T, Miyahara S, Miyake A, Kuwahara H and Furukawa N, Magnetic control of transverse electric polarization in BiFeO3, Nat. Commun. 6 (2015) 1–5.

[10] S. Lee, M.T. Fernandez-Diaz, H. Kimura, Y. Noda, D.T. Adroja, S. Lee, J. Park, V. Kiryukhin, S.W. Cheong, M. Mostovoy, J.G. Park, Negative magnetostrictive magnetoelectric coupling of BiFeO3, Phys. Rev. B - Condens. Matter Mater. Phys. 88 (2013) 1-10.

[11] Tokunaga M, Azuma M and Shimakawa Y, High-Field Study of Strong Magnetoelectric Coupling in Single-Domain Crystals of BiFeO3, J. Phys. Soc. Japan 79 (2010) 064713.

[12] Carranza-Celis D, Cardona-Rodríguez A, Narváez J, Moscoso-Londono O, Muraca D,

Knobel M, Ornelas-Soto N, Reiber A and Ramírez J G, Control of Multiferroic properties in BiFeO3 nanoparticles, Sci. Rep. 9 (2019) 1–9.

[13]   T. Park, G.C. Papaefthymiou, A.J. Viescas, A.R. Moodenbaugh, S.S. Wong, Size-Dependent Magnetic Properties of Nanoparticles, Nano Letters 7 (2019) 766-772.

[14]   Huang F, Wang Z, Lu X, Zhang J, Min K, Lin W, Ti R, Xu T, He J, Yue C and Zhu J, Peculiar magnetism of BiFeO3 nanoparticles with size approaching the period of the spiral spin structure, Sci. Rep. 3 (2013) 1–7.

[15]   Béa H, Bibes M, Fusil S, Bouzehouane K, Jacquet E, Rode K, Bencok P and Barthélémy A, Investigation on the origin of the magnetic moment of BiFeO3 thin films by advanced x-ray characterizations, Phys. Rev. B - Condens. Matter Mater. Phys. 74 (2006) 3–6.

[16]   Bharti V and Zhang Q M, Dielectric study of the relaxor ferroelectric poly(vinylidene fluoride-trifluoroethylene) copolymer system, Phys. Rev. B - Condens. Matter Mater. Phys. 63 (2001) 1–6.

[17]   Furukawa T, Structure and functional properties of ferroelectric polymers, Adv. Colloid Interface Sci. 71–72 (1997) 183–208.

[18]   Maciel M M, Ribeiro S, Ribeiro C, Francesko A, Maceiras A, Vilas J L and Lanceros-Méndez S, Relation between fiber orientation and mechanical properties of nano-engineered poly(vinylidene fluoride) electrospun composite fiber mats, Compos. Part B Eng. 139 (2018) 146–54.

[19]   Martins P, Kolen'Ko Y V., Rivas J and Lanceros-Mendez S, Tailored Magnetic and Magnetoelectric Responses of Polymer-Based Composites, ACS Appl. Mater. Interfaces 7 (2015) 15017–22.

[20]   Pradhan D K, Kumari S and Rack P D, Magnetoelectric composites: Applications, coupling mechanisms, and future directions, Nanomaterials 10 (2020) 1–22.

[21]   Pradhan D K, Puli V S, Kumari S, Sahoo S, Das P T, Pradhan K, Pradhan D K, Scott J

F and Katiyar R S, Studies of phase transitions and magnetoelectric coupling in PFN-CZFO multiferroic composites, J. Phys. Chem. C 120 (2016) 1936–44.

[22] Hu J M, Duan C G, Nan C W and Chen L Q, Understanding and designing magnetoelectric heterostructures guided by computation: Progresses, remaining questions, and perspectives, npj Comput. Mater. 3 (2017) 1–20.

[23] Rotter M, Wang Z S, Boothroyd A T, Prabhakaran D, Tanaka A and Doerr M, Mechanism of spin crossover in LaCoO3 resolved by shape magnetostriction in pulsed magnetic fields, Sci. Rep. 4 (2014) 1–4

[24] D. Hunter, W. Osborn, K. Wang, N. Kazantseva, J. Hattrick-Simpers, R. Suchoski, R. Takahashi, M.L. Young, A. Mehta, L.A. Bendersky, S.E. Lofland, M. Wuttig, I. Takeuchi, Giant magnetostriction in annealed Co1-xFex thin-films, Nat. Commun.2 (2011) 1-7.

[25] Wang H, Zhang Y N, Wu R Q, Sun L Z, Xu D S and Zhang Z D, Understanding strong magnetostriction in Fe 100-x Gax alloys, Sci. Rep. 3 (2013) 1–5.

[26] G. Vertsioti, M. Pissas, S.J. Zhang, D. Stamopoulos, Electric-field control of the remanent-magnetic-state relaxation in a piezoelectric-ferromagnetic PZT-5%Fe3O4 composite, J. Appl. Phys. 126 (2019) 044104.

[27] E. Pan, X. Wang, R. Wang, Enhancement of magnetoelectric effect in multiferroic fibrous nanocomposites via size-dependent material properties, Appl. Phys. Lett .95 (2009) 181904.

[28] A.J. Swift, Surface and interface analysis, Surf. Eng. Caseb. (1996) 255–279.

[29] Ribeiro C, Costa C M, Correia D M, Nunes-Pereira J, Oliveira J, Martins P, Gonçalves R, Cardoso V F and Lanceros-Méndez S, Electroactive poly(vinylidene fluoride)-based structures for advanced applications, Nat. Protoc. 13 (2018) 681–704.

[30] S. Svirskas, M. Simenas, J. Banys, P. Martins, S. Lanceros-Mendez, Dielectric

relaxation and ferromagnetic resonance in magnetoelectric (Polyvinylidene-fluoride)/ferrite composites, J. Polym. Res. 22 (2015) 1-10.

[31] Brito-pereira R, Ribeiro C, Lanceros-mendez S and Martins P, Magnetoelectric response on Terfenol-D / P ( VDF-TrFE ) two-phase composites, Compos. Part B 120 (2017) 97–102.

[32] Gonçalves R, Martins P M, Caparrós C, Martins P, Benelmekki M, Botelho G, Lanceros-Mendez S, Lasheras A, Gutiérrez J and Barandiarán J M, Nucleation of the electroactive β-phase, dielectric and magnetic response of poly(vinylidene fluoride) composites with Fe2O 3 nanoparticles, J. Non. Cryst. Solids 361 (2013) 93–9.

[33] Behera C, Choudhary R N P and Das P R, Development of Multiferroism in PVDF with CoFe2O4 Nanoparticles, J. Polym. Res. 24 (2017) 1-1.3

[34] Bouaamlat H, Hadi N, Belghiti N, Sadki H, Bennani M N, Abdi F, Lamcharfi T, Bouachrine M and Abarkan M, Dielectric Properties , AC Conductivity , and Electric Modulus Analysis of Bulk Ethylcarbazole-Terphenyl 2020 (2020) 1-8.

[35] D.Y. Kusuma, C.A. Nguyen, P.S. Lee, Enhanced ferroelectric switching characteristics of P(VDF-TrFE) for organic memory devices, J. Phys. Chem. B. 114 (2010) 13289–13293.

[36] Szafraniak-Wiza I, Andrzejewski B and Hilczerb B, Magnetic properties of bismuth ferrite nanopowder obtained by mechanochemical synthesis, Proc. 8th Int. Conf. Mechanochemistry Mech. Alloy. INCOME 2014 (2014) 1029–31.

[37] J. Maniks, R. Zabels, R.M. Meri, J. Zicans, Structure, micromechanical and magnetic properties of polycarbonate nanocomposites, IOP Conf. Ser. Mater. Sci. Eng.49 (2013) 012012.

[38] I.J. Bruvera, P. Mendoza Zélis, M. Pilar Calatayud, G.F. Goya, F.H. Sánchez, Determination of the blocking temperature of magnetic nanoparticles: The good, the

bad, and the ugly, J. Appl. Phys.118 (2015) 184304.

[39]   Knobel M, Nunes W C, Brandl A L, Vargas J M, Socolovsky L M and Zanchet D, Interaction effects in magnetic granular systems, Phys. B Condens. Matter 354 (2004) 80–7.

[40]   Guduri B R and Luyt A S, Structure and mechanical properties of polycarbonate modified clay nanocomposites, J. Nanosci. Nanotechnol. 8 (2008) 1880–5.

[41]   Bhoi K, Dash S, Dugu S, Pradhan D K, Rahaman M M, Simhachalam N B, Singh A K, Vishwakarma P N, Katiyar R S and Pradhan D K , Phase transitions and magneto-electric properties of 70 wt. % Pb(Fe0.5Nb0.5)O3-30 wt. % Co0.6Zn0.4Fe1.7Mn0.3O4 multiferroic composite, J. Appl. Phys. 130 (2021) 0–15.

[42]   Chu B, Lin M, Neese B, Zhou X, Chen Q and Zhang Q M, Large enhancement in polarization response and energy density of poly(vinylidene fluoride-trifluoroethylene-chlorofluoroethylene) by interface effect in nanocomposites, Appl. Phys. Lett. 91 (2007) 1–4.

[43]   Martins P, Lasheras A, Gutierrez J, Barandiaran J M, Orue I and Lanceros-Mendez S, Optimizing piezoelectric and magnetoelectric responses on CoFe 2O4/P(VDF-TrFE) nanocomposites, J. Phys. D. Appl. Phys. (2011) 44.

[44]   Kumar A, Patel P K, Yadav K L, Singh Y and Kumar N, Enhanced magnetoelectric coupling response in hot pressed BiFeO 3 and polymer composite films : Effect of magnetic field on grain boundary and grain resistance, Mater. Res. Bull. 145 ( 2022) 111527.

[45]   Feng R, Zhu Z, Liu Y, Song S, Zhang Y, Yuan Y, Han T, Xiong C and Dong L, Magnetoelectric effect in flexible nanocomposite films based on size-matching, Nanoscale 13 (2021) 4177–87.

[46]   Martins P, Moya X, Phillips L C, Kar-Narayan S, Mathur N D and Lanceros-Mendez

S, Linear anhysteretic direct magnetoelectric effect in Ni 0.5Zn0.5Fe2O4/poly(vinylidene fluoride-trifluoroethylene) 0-3 nanocomposites, J. Phys. D. Appl. Phys. 44 (2011) 482001.

[47]   Ahlawat A, Satapathy S, Choudhary R J, Gupta P K, Satapathy S, Choudhary R J, Singh M K and Gupta P K, Tunable room temperature magnetoelectric response of SmFeO3/poly (vinylidene fluoride) nanocomposite films, RSC Advances 6 (2016) 44843-44850.

[48]   Ahlawat A, Satapathy S, Deshmukh P, Shirolkar M M and Sinha A K, Electric field poling induced self-biased converse magnetoelectric response in PMN-PT / NiFe 2 O 4 nanocomposites, Appl. Phys. Lett. 111 (2017) 0–5.

[49]   Geprägs S, Brandlmaier A, Opel M, Gross R, Goennenwein S T B, Geprägs S, Brandlmaier A, Opel M, Gross R and Goennenwein S T B, Electric field controlled manipulation of the magnetization in Ni / BaTiO 3 hybrid structures Electric field controlled manipulation of the magnetization in Ni / BaTiO3, Appl. Phys. Lett. 96 (2011) 3–6.